\documentstyle[amsmath,multicol,epsfig,prb,aps,color,graphicx]{revtex}

\begin{document}
\newcommand{\Eq}[1]{Eq.~(\ref{#1})}
\newcommand{\fg}[1]{Fig.~\ref{#1}}
\newcommand{\cph}[1]{Chem. Phys. {\bf #1}}
\newcommand{\prevb}[1]{Phys. Rev. B{\bf {#1}}}
\newcommand{\prevl}[1]{Phys. Rev. Lett. {\bf {#1}}}
\newcommand{\sm}[1]{Synth. Metals {\bf {#1}}}
\newcommand{\jchemp}[1]{J. Chem. Phys. {\bf {#1}}}
\newcommand{\jphysc}[1]{J. Phys. Chem. {\bf {#1}}}
\newcommand{\jpcb}[1]{{J. Phys. Chem. B {\bf #1}}}
\newcommand{\pcp}{$\pi$-conjugated polymer}
\newcommand{\red}{\color{red}}
\newcommand{\blue}{\color{blue}}
\newcommand{\w}{\mbox{$\omega$}}
\newcommand{\D}{\mbox{$\Delta$}}
\newcommand{\G}{\mbox{$\Gamma$}}
\newcommand{\smq}{\mbox{$\simeq$}}
\newcommand{\ee}{E. Ehrenfreund}
\newcommand{\zvv}{Z.V. Vardeny}
\newcommand{\bi}{\bibitem}
\newcommand{\beq}{\begin{equation}}
\newcommand{\eeq}{\end{equation}}
\newcommand{\technion}{Technion--Israel Institute of Technology,
Haifa 32000, Israel}
\newcommand{\ssi}{Solid State Institute}
\newcommand{\phy}{Department of Physics}
\newcommand{\hs}{\hspace{0.4cm}} %standard horizontal space
\newcommand{\npi}{\hspace{-0.4cm}} %no paragraph indent
\newcommand{\vs}{\vspace{0.4cm}} %standard vertical space

\newcounter{fg}
 \refstepcounter{fg} \label{PL} \refstepcounter{fg} \label{Raman}
\refstepcounter{fg} \label{model}

\title{
 Apparent phonon side band modes in $\pi$-conjugated
systems: polymers, oligomers and crystals}

\author{
 \ee$^{1}$, C.C. Wu$^2$, \zvv$^{2}$}
\address{$^1$\phy, \technion\\
$^2$\phy, University of Utah, Salt Lake City, UT 84112
 }

\maketitle

\begin{abstract}

The emission spectra of many  $\pi$-conjugated polymers and
oligomers contain side-band replicas with apparent frequencies
that do not match the Raman active mode frequencies. Using a time
dependent model we show that in such many mode systems, the
increased damping of the time dependent transition dipole moment
correlation function results in an effective elimination of the
vibrational modes from the emission spectrum; subsequently causing
the appearance of a regularly spaced progression at a new apparent
frequency. We use this damping dependent vibrational reshaping to
quantitatively account for the vibronic structure in the emission
spectra of  $\pi$-conjugated systems in the form of films, dilute
solutions and single crystals. In particular, we show that by
using the experimentally measured Raman spectrum we can account in
detail for the apparent progression frequencies and their relative
intensities in the emission spectrum.

\vs

\npi {\bf Keywords:} Optical absorption and emission spectroscopy,
Photoluminescence, Raman spectroscopy, Conjugated and/or
conducting polymers

\end{abstract}

\setlength{\columnsep}{6mm}
\begin{multicols}{2}%this line for double column version
%\section
\npi {\bf 1. Introduction}

   The typical photoluminescence (PL)
spectrum of many $\pi$-conjugated systems is composed of vibronic
progression series (e.g., Fig. 1a, [1]). One key feature of the
vibronic progression is that it contains side-band replicas with
apparent frequencies that do not match the measured Raman active
modes (e.g., Fig. 1b, [1]). In many cases the rich Raman spectrum
is reduced to only one or two vibronic progression in the emission
spectrum, not necessarily having the same frequencies as the most
intense Raman modes. In this work we account for the apparent
modes that appear in the PL spectra of  $\pi$-conjugated systems,
such as polymers, oligomers and single crystals. We use the
measured pre-resonance Raman spectra in order to quantify the
relative configuration displacement (or, equivalently, the
Huang-Rhys factor) for each of the modes. We then utilize these
relative Huang-Rhys (HR) factors in a damped time dependent model
for the PL emission. We show that the apparent vibronic
frequencies in the PL spectrum are the result of a "weighted
beating" of all Raman frequencies.  These frequencies are not, in
general, the simple average of the Raman frequencies, and thus
cannot be predicted a priori. Furthermore, the relative
intensities of the apparent vibronic structure are uniquely
determined by their strengths in the Raman spectrum.

\vs

%\section
\npi {\bf 2.  Time dependent multi-vibrational analysis of
photoluminescence}

In order to account for the PL spectrum of a multi-vibrational
system in relation to its Raman spectrum, it is useful to employ
time dependent analysis rather than the often used sum-over-states
Franck-Condon approach [2]. We write the PL spectrum, F(E), as a
Fourier transform [3]:
 \beq F(E)=\int_{-\infty}^{\infty}f(t)
\textrm{exp}\left(\frac{iEt}{\hbar}\right)dt~. \label{1}
 \eeq
where the dipole correlation function, f(t), is given by:
\begin{equation}
 f(t)=|P|^2\textrm{exp}\left[\frac{-iE_0t}{\hbar}-S+S_+(t)+S_-(t)-\Gamma |t|\right]~.\label{2}
 \end{equation}
In Eq. (2), P is the dipole matrix element and $E_0$ is the bare
energy for the relevant optical transition. $S$ and $S_{\pm}(t)$
are given by the following sums over the vibration modes:
\begin{eqnarray}
  S_{+}(t)&=&\sum_jS_j(n_j+1)\textrm{exp}({-i\w_jt})~; \notag \\
   S_{-}(t)&=&\sum_jS_jn_j\textrm{exp}({i\w_jt})~;\label{3}  \\
S&=&\sum_jS_j(2n_j+1)~,\notag
 \end{eqnarray}
where  $\w_j$ are the Raman mode frequencies, $n_j$ are the
equilibrium mode populations according to the Bose-Einstein
distribution, $S_j=\w_j\D_j^2/2\hbar$ ($\D_j$ is the mode
equilibrium displacement in the optically excited electronic state
relative to the ground state) are identified as the mode HR
factors, whereas S=S(T) is the total temperature dependent HR
factor. We emphasize here that it is the "electron temperature",
$T_e$, which determines the mode occupation, $n_j$. $T_e$ is
determined by the photon excitation energy and the electron excess
energy relaxation rate and may be considerably higher than the
lattice temperature, T. The time dependent term $S_+(t)$ is
responsible for the red shifted vibronic side bands in the PL
spectrum, while the $S_-(t)$ term gives rise to blue shifted (due
to a "hot luminescence" process) side bands, which may appear for
low frequency modes at relatively high temperatures.  In Eq. (2)
we introduced a simple mode independent phenomenological damping
parameter $\G$; it represents losses due to natural line
broadening and/or other degrees of freedom [4].

\begin{figure}
\begin{center}
\includegraphics[width=7.cm]{Fig1.prn}
\end{center}
\end{figure}
\vspace{-0.9cm}
\begin{figure}
\begin{center}
\includegraphics[width=6.cm,angle=-90]{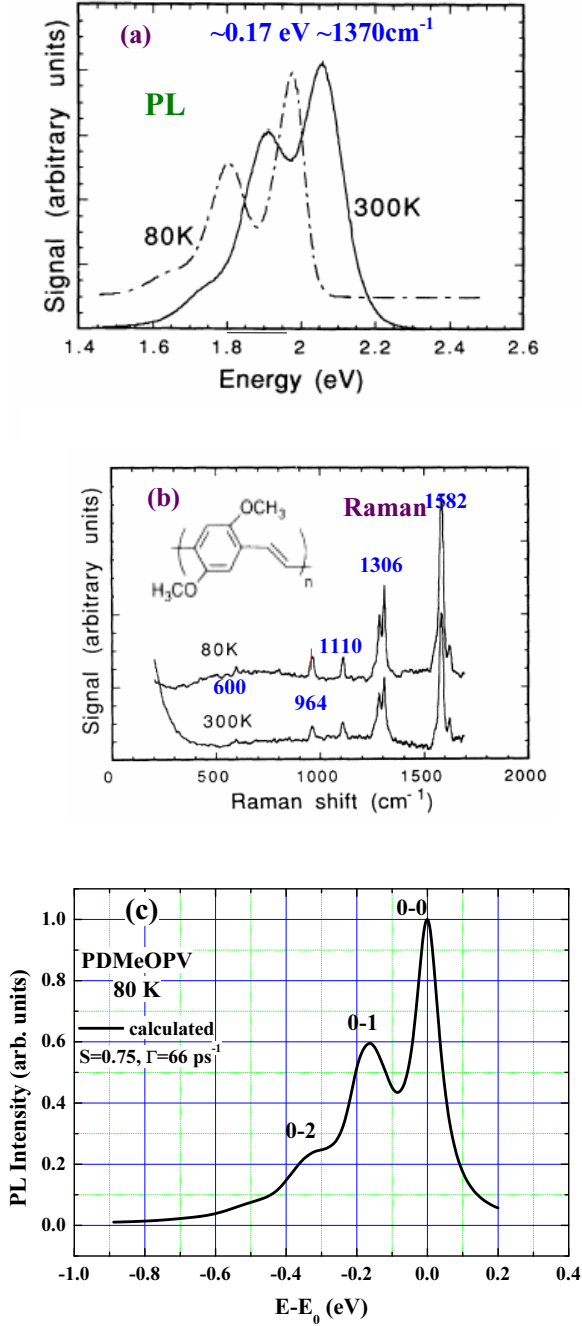}
\caption{ (a) PL spectrum of PDMeOPV; showing a single vibronic
frequency at $\approx$1370 $cm^{-1}$. (b) Raman spectrum of
PDMeOPV. The Raman frequencies are marked. Inset: chemical
structure. (Data from: Woo et. al., Ref. [1]). (c) Calculated PL
spectrum using the data of Fig. 1b and Eqs. 1-3. The resulting
vibronic separation is $\approx$1350 $cm^{-1}$. Energy is measured
relative to the 0-0 transition energy ($E_0$).}
\end{center}
\end{figure}

The HR factors that determine the vibronic structure are closely
related to the measured Raman spectrum, since the Raman process is
enabled by the same electron-phonon interaction.  The T=0 K
intensity, $I_j^0$, of each Raman line measures its excited state
displacement  $\D_j$ [2]: $I_j^0\propto\w_j^2$; we then have:
$S_j\propto I_j^0/\w_j^2$. We note the proportionality of $S_j$ to
$\w_j^{-2}$, which emphasizes the contribution of lower frequency
Raman modes to the vibronic structure in the PL emission.

As an example for the above analysis, we consider films of
poly(dimethoxy phenylene vinylene) [PDMeOPV]. Using the Raman data
published previously [1] (Fig. 1b), we have calculated the
expected PL vibronic structure. This is shown in Fig. 1c. As seen,
the calculated spectrum is in agreement with the experimental
spectrum: (a) Both spectra are composed of a single vibronic
progression; (b) The calculated progression frequency of $\approx$
0.16 eV in the PL spectrum (Fig. 1c) is very close to the measured
one (Fig. 1a). A choice of total HR factor of S(0)=0.75 and
damping of $\G $=66 ps$^{-1}$ yields relative intensities and line
broadening similar to that measured at T=80 K.

\vs
%\section
\npi {\bf 3. The case of distyryl-benzene}

The excited state properties of distyryl-benzene (DSB), which is
the three phenyl group oligomer of p-phenylene vinylene (Fig. 2a,
inset), have been the subject of recent experimental [5,6] and
theoretical [7] spectroscopic studies, because of potential
opto-electronic applications [8]. DSB PL spectroscopy has been the
subject of numerous research studies, since it is strongly
dependent upon the packing order. The first excited state of
isolated DSB molecules is optically allowed, making them strongly
luminescent and useable as active media in light emitting devices
[8]. The room temperature PL spectrum of isolated DSB molecules
consists of the fundamental ("0-0") optical transition and a
single apparent frequency phonon side band replica series. In DSB
films the molecules form H aggregates, thus substantially
weakening the PL emission quantum yield relative to the isolated
oligomers in dilute solutions [9]. This is especially true in DSB
single crystals: due to the herring bone symmetry the fundamental
optical transition is either totally absent or significantly
reduced [9]. Typical PL spectra of DSB chromophores contain
vibronic progression series, of which the relative intensity and
frequencies depend on the packing and temperature. The PL emission
spectra of crystalline DSB are discussed elsewhere [10]. Here, we
discuss only our results for dilute solutions, which reflect the
single molecule PL spectrum.

\vs

%\subsection
\npi {3.1. DSB solution}

In Table I we list the most intense Raman modes observed   for
DSB. The Raman spectra for the solution or crystalline forms are
similar. Here we present the crystalline data since its signal to
noise ratio is much better. Along with the frequencies and
relative intensities,
 we list for each mode the deduced relative HR factor, calculated using
 the model discussed in Section 2 and the Raman data presented in Fig. 2.
 It is important to recognize that the lowest
 frequency mode has the largest HR factor, although its intensity is only
  $\approx$1\% of the strongest line!

\begin{figure}
\begin{center}
\includegraphics[width=8.cm,angle=0]{Fig2.prn}
\caption{(a) PL spectroscopy of DSB in solution at 20 and 200 K.
The indices 0 and 1,2,3 in the 200 K spectrum denote the
fundamental and vibronic replica transitions, respectively. The
pairs of indices k, n (k=0,1,2,3) in the 20 K spectrum denote the
complex modulated vibronic structure (see text). Inset: Chemical
structure of DSB molecule. (b) High sensitivity preresonant Raman
spectrum of DSB crystal at 10 K. The Raman frequencies and
intensities are listed in Table 1.}
\end{center}
\end{figure}

\vspace{-0.2cm}

In Fig. 2a we show the PL spectra of DSB in dilute frozen solution
of tetradecane [11] at low and high temperatures.  The measured PL
spectrum at T=200 K has the typical vibronic progression shape as
other  $\pi$-conjugated systems, but notable changes occur with
decreasing temperature. At T=200 K (Fig. 2a, bottom curve), there
appears a dominant "high frequency" vibronic progression of
$\approx$0.17 eV ($\approx$1370 $cm^{-1}$), which, however does
not match any of the Raman frequencies (Fig. 2b); it is not even
in the close vicinity of the strongly coupled modes (Table I). The
highest energy PL peak (marked "0") is interpreted as the
fundamental optical transition ($1A_g\rightarrow 1B_u$), whereas
the lower energy peaks (marked "1,2,3") are the vibronic side
bands. At low temperatures, the high frequency progression is
modulated by a different "low frequency" progression of
$\approx$17-19 meV (Fig. 1a, T=20 K). We denote this modulated
vibronic structure by k, n (k=0,1,2,3, n=0,1,2,...), where k (n)
is the order of the high (low) frequency modulation.

Using the data of Table I we show in Fig. 3a the dipole
auto-correlation function, f(t), generated for low and high
damping. It is visually striking that the 11 mode system is
dominated by only two apparent modes: a short period mode
modulated by a long period mode. Moreover, the frequencies
associated with these two modes do not coincide with any DSB
normal mode. The data given in Table I is used also to generate
the PL spectra shown in Fig. 3b. Here, the values of the electron
temperature, $T_e$, HR factor S and damping $\G$  were adjusted to
best fit the frozen solution experimental data at low and high
lattice temperatures, T (Fig. 2a).

\vspace{0.1cm}
%\newpage
\[
\begin{array}{c|cccccc} \hline \hline
~j&1&2&3&4&5&6\\
 ~\nu~(cm^{-1}) &131&261&640 &873&1000 &1181 \\
 ~I/I_{10}(\%)&1.2&0.45&0.71 &0.77&6.7 &50 \\
~S&0.88&0.098&0.025 &0.025&0.088 &0.48 \\ \hline
 ~j&7&8&9&10&11&~\\
~\nu~(cm^{-1}) &1330&1452&1561 &1591&1635& \\
 ~I/I_{10}(\%)&19&2.7&13 &100&42& \\
~S&0.15&0.018&0.068 &0.50&0.20& \\
\hline \hline
\end{array} \]

Table I. {\small The most intense Raman lines of DSB at T=10 K.
$\nu$, $I$ and $S$ are the frequency, relative intensity and HR
factor, respectively, for each mode.}

\vspace{5mm}

 The higher damping spectrum that presumably occurs at
T=200 K ($\G$=42 ps$^{-1}$, Fig. 3b) shows a vibronic progression
dominated by a single frequency of 173 meV ($\approx$1400
$cm^{-1}$). These progression peaks are marked as "1,2,3". This is
in excellent agreement with the experimental data at T=200 K (Fig.
2a, bottom curve). For this case we chose $T_e=T$, since the
relevant modes are at high frequencies. At lower lattice
temperatures we expect the damping to decrease. Consequently, the
low frequency apparent mode is less suppressed, making the PL
spectrum sensitive to the actual value of $T_e$. For low $\G$
values, there appears a low frequency modulation ($\approx$17 meV)
of the high frequency vibronic series, as seen in Fig. 3b (top
curve). The combined progression peaks are denoted as (k,n), where
k=0,1,2,3 denotes the high frequency apparent progression and
n=0,1,2... denotes the low frequency modulation. The PL spectrum
in Fig. 3b (top curve) was calculated using $T_e$=150 K; it shows
two blue shifted peaks, in very good agreement with the
experimental data at T=20 K (Fig. 2a, top curve). We thus conclude
that due to the non-resonant PL excitation the electron
temperature is much higher than the lattice temperature: $T_e>T$.

\vspace{-3mm}

\begin{figure}
\begin{center}
\includegraphics[width=8.cm,angle=0]{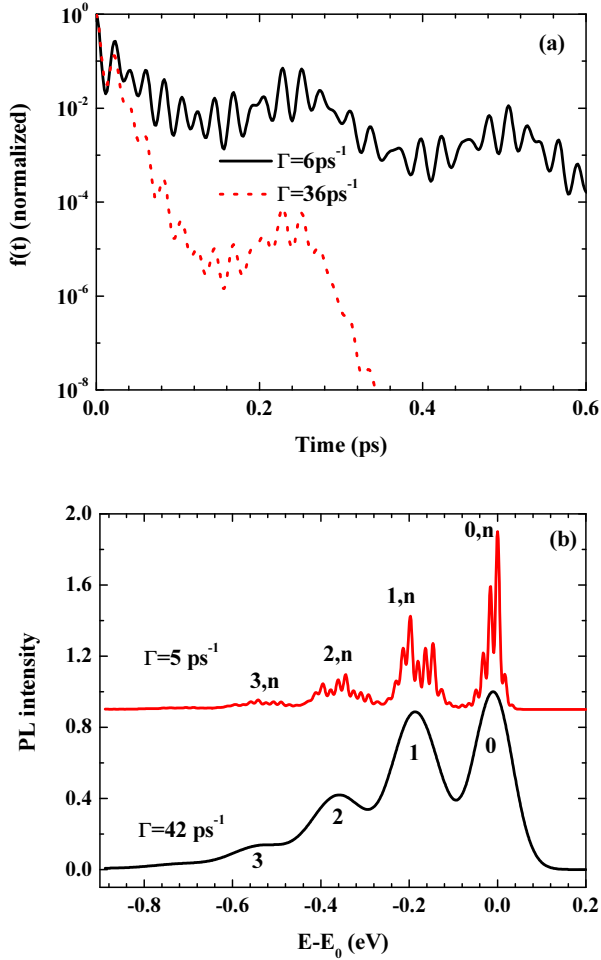}
\caption{Model calculations of PL spectra in DSB using the 11 mode
system of Table I. (a) Normalized dipole auto-correlation function
f(t) [Eqs. (2),(3)],  for HR factor S(T=0)=2.5 and two values for
$\G$, as shown. Note the log scale. (b) PL spectra obtained by the
Fourier transform of f(t) similar to that in (a), but for
S(0)=1.7, $T_e$=150 (200) K for the upper (lower) curve and $\G$
as shown. The energy is measured relative to the fundamental
transition.}
\end{center}
\end{figure}

%\vspace{-7mm}

In summary, we have shown that increased damping in
multi-vibrational  $\pi$-conjugated systems results in effective
elimination of vibrational modes from the emission and absorption
spectra and the eventual appearance of a nearly regularly spaced
progression at an apparent frequency. We have presented a method
by which measuring the full Raman spectrum, the emission spectrum
of  $\pi$-conjugated systems can be account for in detail. In
particular, we interpret the blue shifted small peaks above the
0-0 transition, in the frozen DSB solution spectrum, as due to a
"hot luminescence" process. In such a process, due to a high
electron temperature, the first excited vibrational level is
significantly populated.

{\bf Acknowledgments}--Supported in part by DOE grant FG-04
ER46109 and by the Israel Science Foundation 735/04.

\end{multicols}
\end{document}